\documentclass[numberedrefs,preprint,turnofflinenumbers]{JASA-EL}

\begin{document}

\title{Passive ladder-network representation of the radiation impedance of a spherical cap on a rigid sphere}

\author{Ronald M. Aarts}
\email{r.m.aarts@tue.nl}
\correspondingauthor
\affiliation{Department of Electrical Engineering, Eindhoven University of Technology, Eindhoven 5600 MB, The Netherlands}

\date{\today}

\begin{abstract}
\noindent A loudspeaker in a closed box is modeled as a rigid sphere with an axially
moving cap. For an axisymmetric cone-velocity profile, its radiation impedance is
written as a network. Each spherical-harmonic mode has an impedance rational in
frequency, realized exactly as a finite passive ladder of masses and compliances terminated in
a resistance. Ideal transformers set by Legendre coefficients couple these ladders; truncated modes
reduce to a residual mass. Despite the different geometry, the impedance resembles that of a
baffled piston but, at equal volume velocity, has half its low-frequency resistance and about
three quarters of its radiation mass.
\end{abstract}

\maketitle

\section{Introduction}
The rigid circular piston in an infinite baffle is the standard radiation model for a
loudspeaker, and its radiation impedance is known in closed form in terms of Bessel and
Struve functions.\citep{beranek2012,aj2003,aj2016} For a driver in a closed box, a rigid
sphere with a moving spherical cap is more realistic.\citep{morse1968} With comparable box and
sphere volumes and equal membrane and cap areas, its polar response, baffle step, sound
power and directivity are close to those of a real loudspeaker, and its polar response is
closer than that of the baffled piston.\citep{aj2010,aj2011} Radiation from caps on spheres has
a long history. Morse and Ingard treated the radially vibrating cap,\citep{morse1968} and Sherman the self- and
mutual radiation impedances of sources on a sphere.\citep{sherman1959} Recently, Bellows and Leishman used the
modal self-impedance of a radially vibrating cap and reduced it at low frequency to an
exterior end correction, i.e., a single mass,\citep{bellows2023} which Bellows compared
with those of baffled and unbaffled circular radiators.\citep{bellows2024} Closest in aim, H\'elie
and Rodet averaged the impedance of a uniformly pulsating cap over its surface and fitted
low-order parametric models with checked stability for real-time simulation.\citep{helie2003} For
lumped-element loudspeaker design, however, the radiation load is wanted as a circuit that
is passive, so that it is stable in time-domain simulation, and accurate over a prescribed
bandwidth, not only at low frequency.

This Letter gives such a circuit for a cap that moves \emph{axially}, as a loudspeaker cone
does, with an arbitrary axisymmetric velocity profile. The modal series for the impedance is
standard; the present contribution is its exact network realization. It consists of (i) an
exact finite Cauer ladder for each spherical mode, with explicit positive, frequency-independent
element values in closed form; (ii) the coupling of these ladders to the axially moving cap through passive
ideal transformers; and (iii) a residual mass that turns the truncated modal series into a
finite network of controllable accuracy. The ingredients are known: the modal expansion on the
sphere,\citep{morse1968,aj2010} the rational form of the modal impedance, and its
continued-fraction (Cauer) synthesis.\citep{guillemin1957} The latter two steps are the
acoustic analogue of Chu's circuit synthesis for spherical electromagnetic
modes.\citep{chu1948,hansen1981} Continued fractions and passive ladders for acoustic
impedances that are not rational, such as the baffled piston and the horn, are treated in a
companion paper.\citep{aarts2026} Network representations of spherical
sources of arbitrary order were also reported early on by Kjerbye Nielsen;\citep{kjerbye1956} only a
short meeting abstract is available, which gives no details of the networks. Unlike Ref.~\citenum{helie2003},
the network here is exact per mode rather than fitted, and it represents an axially moving cap
with an arbitrary profile rather than a uniformly pulsating one.

\section{Modal radiation impedance as a finite ladder}
Consider a rigid sphere of radius $R$ in a fluid of density $\rho$ and sound speed $c$, with
time dependence $e^{j\omega t}$, wavenumber $k=\omega/c$, and polar angle $\theta$ measured from
the axis. A normal surface velocity $V_nP_n(\cos\theta)$, with $P_n$ the Legendre polynomial of
degree $n=0,1,2,\ldots$ and $V_n$ a complex amplitude, produces a pressure on the sphere
$Z_nV_nP_n(\cos\theta)$, with the specific modal impedance\citep{morse1968}
\begin{equation}
Z_n=-j\rho c\,\frac{h_n^{(2)}(kR)}{h_n^{(2)\prime}(kR)}
=\rho c\,\frac{s\,q_n(s)}{(s+n+1)\,q_n(s)-s\,q_n'(s)},\qquad s=jkR .
\label{eq:Zn}
\end{equation}
Here $q_n(s)=\sum_{m=0}^{n}\frac{(n+m)!}{(n-m)!\,m!\,2^m}\,s^{\,n-m}$ is the reverse
Bessel polynomial, which follows from
$h_n^{(2)}(\xi)\propto e^{-j\xi}q_n(j\xi)/\xi^{n+1}$ (Ref.~\citenum{olver2010}, Sec.~10.49). For example, $Z_0=\rho c\,s/(s+1)$ and
$Z_1=\rho c\,s(s+1)/(s^2+2s+2)$.

Because $Z_n/\rho c$ is a positive-real rational function, alternately removing poles at
$s=0$ from admittance and impedance (Cauer's second form) gives a high-pass ladder that
terminates in $\rho c$. The element values follow from the recurrence
$h_{m-1}^{(2)}+h_{m+1}^{(2)}=(2m+1)h_m^{(2)}/\xi$, exactly as in Chu's derivation for
electromagnetic modes:\citep{chu1948,hansen1981} with $\xi=kR$,
\[
\frac{\rho c}{Z_n}=\frac{n+1}{s}+j\frac{h_{n-1}^{(2)}}{h_n^{(2)}},\qquad
-j\frac{h_m^{(2)}}{h_{m-1}^{(2)}}=\frac{2m-1}{s}+j\frac{h_{m-2}^{(2)}}{h_{m-1}^{(2)}},\qquad
j\frac{h_{-1}^{(2)}}{h_0^{(2)}}=1,
\]
so that a continued fraction with partial denominators $(n+1)/s,(2n-1)/s,(2n-3)/s,\ldots,1/s$,
closed by 1, results. All these coefficients are positive, and each term $c/s$ with $c>0$
is a positive mass (in shunt) or compliance (in series); together with the positive termination
this makes the ladder passive. Positivity is easy to establish here because $Z_n$ is rational;
for the non-rational impedance of the baffled piston,\citep{aarts2026} the positivity of the
continued-fraction coefficients is harder to prove and is shown in Ref.~\citenum{aartsproof}. In physical units the elements are
\begin{equation}
M_0=\frac{\rho R}{n+1},\qquad
C_1=\frac{R}{(2n-1)\rho c^2},\quad M_2=\frac{\rho R}{2n-3},\quad
C_3=\frac{R}{(2n-5)\rho c^2},\ \ldots
\label{eq:elements}
\end{equation}
These values were also confirmed symbolically for $n\le8$ and numerically for $n\le30$
(relative error $<10^{-14}$). Each modal ladder is a nested structure, the circuit form of the
continued fraction. The ladders of different modes are not nested in one another: all element
values depend on $n$, so the ladder of mode $n+1$ does not contain that of mode $n$, and the modes
are connected in series instead (Sec.~3). The ladder has $n+1$ reactive elements: masses $M$ in shunt and compliances $C$ in series,
counted from the input. The coefficients $2n-1,2n-3,\ldots,1$ run down to 1, after which
the ladder is closed by $\rho c$. All elements are specific (per unit area), positive, and
independent of frequency. For $n=0$ and $n=1$ these are the textbook circuits of the pulsating
and the oscillating sphere;\citep{beranek2012} for $n=0$, a mass $\rho R$ in parallel with $\rho c$. The input mass $\rho R/(n+1)$ is the low-frequency entrained mass of
mode $n$. For $kR\ll n$ the mode is almost purely reactive, $Z_n\approx j\omega\rho R/(n+1)$.

\section{Axially moving cap}
\begin{figure}[ht]\centering
\includegraphics[width=\textwidth]{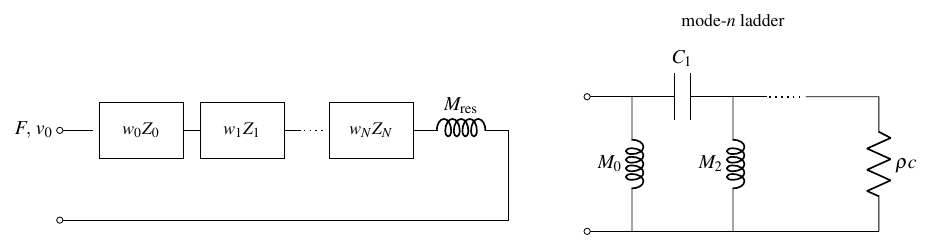}
\caption{Network for the axially moving cap. Left: at the mechanical port $(F,v_0)$ the modal
impedances are connected in series. Each block $w_nZ_n$ is an ideal transformer with impedance
ratio $w_n$ (turns ratio $1:\sqrt{w_n}$), loaded by the ladder of mode $n$, for $n\le N$. The modes above $N$ are combined into the residual mass $M_{\rm res}$. Right:
the ladder of mode $n$ (impedance analogy, mass $\leftrightarrow$ inductor) with the elements
of Eq.~(2). The ladder ends after $n+1$ reactive elements in $\rho c$.}
\label{fig:net}
\end{figure}

Let the cap $0\le\theta\le\theta_0$ move as a loudspeaker cone, parallel to the axis, with
axial velocity $v_0f(\theta)$, with $v_0$ the complex velocity amplitude and $f$ a real,
frequency-independent profile. The sphere elsewhere is at rest. The normal velocity is then
$W(\theta)=v_0f(\theta)\cos\theta$ on the cap and zero elsewhere, as in Ref.~\citenum{aj2010}. Its
Legendre coefficients are
$W_n=\tfrac{2n+1}{2}\int_{x_0}^{1}W(\theta)\,P_n(x)\,\mathrm{d}x$, with $x=\cos\theta$,
$x_0=\cos\theta_0$, and $P_n$ as in Sec.~2. For a rigid cap ($f=1$) they have the closed form
\begin{equation}
\frac{W_n}{v_0}=\frac12\Big[(n+1)I_{n+1}+n\,I_{n-1}\Big],\qquad
I_0=1-x_0,\quad I_l=\frac{P_{l-1}(x_0)-P_{l+1}(x_0)}{2l+1}\ (l\ge1),
\label{eq:Wn}
\end{equation}
which is equivalent to Eq.~(7) of Ref.~\citenum{aj2011}. With $p$ the sound pressure on the sphere and $\mathrm{d}S$ its surface element, define the
generalized force on the cap as $F=\int_{\rm cap}p\,f(\theta)\cos\theta\,\mathrm{d}S$, which for a
rigid cap ($f=1$) is the axial force. Then $Fv_0^*=\oint pW^*\,\mathrm{d}S$, the integral being
over the whole sphere, and by the orthogonality of the Legendre polynomials the modes decouple in
the mechanical radiation impedance:
\begin{equation}
Z_{\rm m}\equiv\frac{F}{v_0}=\sum_{n=0}^{\infty}w_nZ_n,\qquad
w_n=\frac{4\pi R^2}{2n+1}\Big|\frac{W_n}{v_0}\Big|^2 .
\label{eq:Zm}
\end{equation}
The weights $w_n$ are non-negative, so the
ideal transformers that realize them below are passive. The volume velocity is $Q=4\pi R^2W_0=v_0A_f$, which defines the effective
area $A_f$ of the profile; the acoustic impedance is $Z_A=Z_{\rm m}/A_f^2$. For the rigid cap
$A_f=S_d=\pi R^2\sin^2\theta_0$, the projected area, the equivalent of the membrane area of a
driver. Equation~\eqref{eq:Zm} is a series connection, at the mechanical port $(F,v_0)$, of
ideal transformers with impedance ratio $w_n$: transformer $n$ passes the velocity
$v_n=\sqrt{w_n}\,v_0$ to its ladder, whose force $F_n=Z_nv_n$ returns as $\sqrt{w_n}\,F_n$, and
$F=\sum_n\sqrt{w_n}\,F_n$. Each transformer is loaded by the ladder of Eq.~\eqref{eq:elements}; see
Fig.~\ref{fig:net}.

The modes have a direct physical reading. $W_0=Q/(4\pi R^2)$ is the mean normal velocity
over the sphere; the net volume velocity of the cap is not compensated by the resting sphere.
At low frequency the cap therefore radiates as a monopole, as a closed box does. Mode $n=1$
carries the front--back asymmetry, and the transition from radiation into $4\pi$ to radiation
into $2\pi$ (the baffle step) involves the modes up to $kR\approx1$--2. The higher modes
represent the velocity discontinuity at the rim of the cap. They are mainly entrained mass and
start to radiate near $kR\approx n$. For a rigid cap in the limit $\theta_0\to\pi$ only $W_1$
survives, which gives the oscillating sphere.

\emph{Truncation.} Modes $0\ldots N$ are kept exactly. Because $Z_n\approx j\omega\rho R/(n+1)$
for $n\gg kR$, the remaining modes are lumped into one mass,
\begin{equation}
M_{\rm res}=\rho R\sum_{n>N}\frac{w_n}{n+1},
\label{eq:Mres}
\end{equation}
which is placed in series at the mechanical port. The total low-frequency radiation mass is
$M_r=\rho R\sum_{n\ge0}w_n/(n+1)$. For a radially vibrating cap of area
$S_c=2\pi R^2(1-x_0)$, $M_r/S_c^2=\rho l^+/S_c$, with $l^+$ the exterior end correction of
Eq.~(41) of Ref.~\citenum{bellows2023}. The complete (untruncated) network thus realizes the full
modal impedance at all frequencies, and its low-frequency limit is that end correction; a
truncated network approximates it over a bandwidth set by $N$. The network is passive for every $N$.

\section{Results}
All results use $\theta_0=\pi/8$ and $R=8.2$\,cm. In Refs.~\citenum{aj2010,aj2011} this cap
has the curved area of a 3.2\,cm driver; its projected radius is $a_d=R\sin\theta_0=3.14$\,cm,
and the comparisons below use equal projected areas.

\emph{Validation.} The real part of Eq.~\eqref{eq:Zm} reproduces the sound-power series of
Ref.~\citenum{aj2011}. The Wronskian gives $\Re Z_n=\rho c/[(kR)^2|h_n^{(2)\prime}(kR)|^2]$, so
$|v_0|^2\Re Z_{\rm m}$ is identical to their Eq.~(17). Numerically, the ladder network with $N=30$
and the residual mass agrees with that series to a relative error below $10^{-9}$ for
$kR\le20$ and $\theta_0\in\{5\pi/32,\pi/8,\pi/10\}$. Because the residual mass is purely
reactive, this checks the resistance only. The complete complex impedance, including the
reactance, follows in principle from the surface pressure of Refs.~\citenum{aj2010,aj2011}, where
only its real part (the sound power) was evaluated. It was checked here against the
limit of a small cap on a large sphere: at fixed $ka_d$ the rigid cap must approach the baffled
piston as $\theta_0\to0$. For $ka_d=0.5$, 1 and 2, the relative difference
$|Z_A-Z_A^{\rm piston}|/|Z_A^{\rm piston}|$ is 12\,\%, 10\,\% and 10\,\% for $\theta_0=\pi/16$, 6.2\,\%,
5.3\,\% and 4.9\,\% for $\pi/32$, and 3.2\,\%, 2.7\,\% and 2.5\,\% for $\pi/64$, decreasing in proportion
to $\theta_0$.

\begin{figure}[ht]\centering
\includegraphics[width=\linewidth]{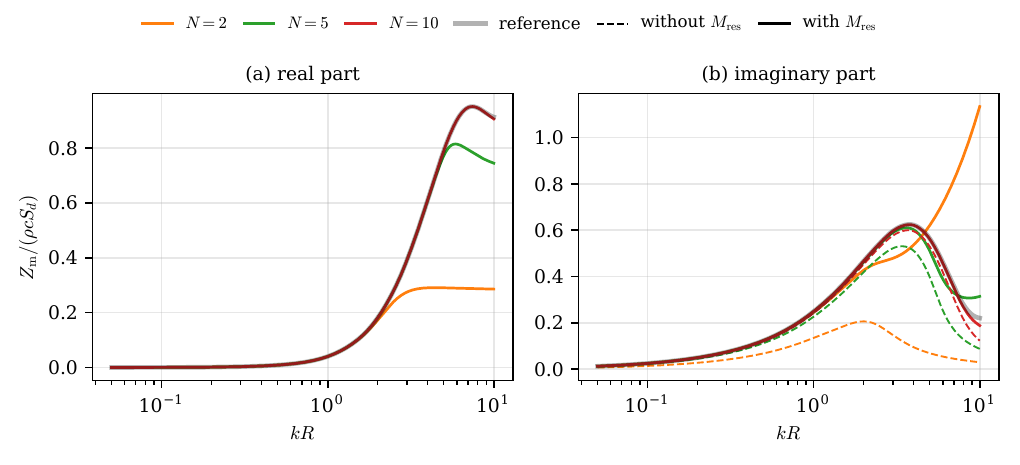}
\caption{Normalized mechanical radiation impedance $Z_{\rm m}/(\rho cS_d)$ of a rigid, axially
moving cap ($\theta_0=\pi/8$), from ladder networks truncated at $N=2,5,10$, compared with the
converged result ($N=300$ plus residual mass, gray). (a) Real part: the residual mass is purely
reactive, so the curves with and without it coincide; for $N=10$ the curve lies on the reference
(within 0.7\,\%). (b) Imaginary part, without (dashed) and with (solid) the residual mass; colors
indicate $N$ in both panels.}
\label{fig:conv}
\end{figure}

\emph{Convergence.} Figure~\ref{fig:conv} shows $Z_{\rm m}/(\rho cS_d)$ for a rigid cap with $N=2,5,10$, with
and without the residual mass. With the error measured as
$|Z_{\rm m}-Z_{\rm m}^{\rm ref}|/|Z_{\rm m}^{\rm ref}|$ and its maximum reached at the upper limit $kR=5$, the
error is 17\,\% for $N=5$ without $M_{\rm res}$ and 5.8\,\% with it. For $N=10$ these errors are
3.6\,\% and 0.2\,\%. Up to $kR=1$, $N=5$ with $M_{\rm res}$ is accurate to better than 0.1\,\%. The residual mass therefore removes most of
the truncation error with the addition of only one reactive element. For the cases
investigated, an empirical rule is $N\gtrsim kR_{\max}+5$. If
$N$ is too small compared with $kR$, the residual mass overestimates the reactance (curve
$N=2$), because the condition $n\gg kR$ behind Eq.~\eqref{eq:Mres} is then violated.

\begin{figure}[ht]\centering
\includegraphics[width=\linewidth]{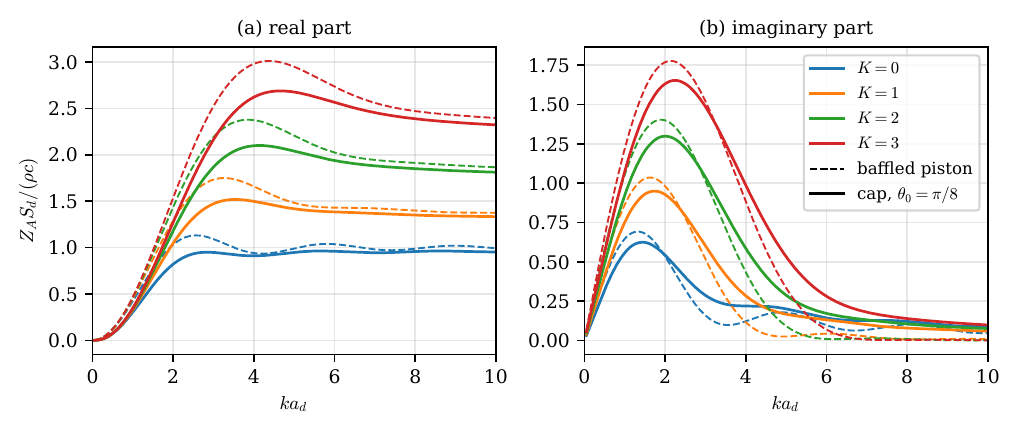}
\caption{Normalized acoustic radiation impedance $Z_AS_d/(\rho c)$ for Stenzel-type profiles
$K=0\ldots3$: baffled piston (dashed, as in Fig.~5 of Ref.~\citenum{aj2009}) and axially moving
cap on a rigid sphere with $\theta_0=\pi/8$, from the ladder network ($N=60$ plus residual
mass; solid). $a_d$ is the piston radius and the projected cap radius.}
\label{fig:piston}
\end{figure}

\emph{Comparison with the baffled piston.} Figure~\ref{fig:piston} compares $Z_AS_d/(\rho c)$ of
the cap with that of a baffled piston of the same projected area. The velocity profiles are of
Stenzel type: $v=(K+1)V_s(1-\sigma^2/a_d^2)^K$ on the piston\citep{aj2009} and
$v_0(K+1)[(\cos\theta-x_0)/(1-x_0)]^K$ on the cap (Eq.~(40) of Ref.~\citenum{aj2010}).
Each impedance is normalized by its own volume velocity. The piston curves are the
quantity $P/(\pi\rho cV_s^2a_d^2)$ of Fig.~5 in Ref.~\citenum{aj2009}; for $K=0$ they agree
with the Struve form to $10^{-7}$. Although the geometries are quite different, the two
impedances are similar in shape. At high frequency, where the sphere is large compared with the
wavelength, the cap behaves much like the baffled piston; at low frequency it shows the
radiation into $4\pi$ of a small closed box. Together with the agreement of the polar response,
baffle step, sound power and directivity with those of a real loudspeaker,\citep{aj2010,aj2011,aj2012}
this supports the cap as a model of the radiation load of a closed-box loudspeaker; a comparison
with measured radiation impedances is outside the scope of this Letter. Three differences stand
out:
\begin{enumerate}
\item For profiles with nonzero net volume velocity, such as these, the low-frequency
radiation resistance of the cap is exactly half that of the piston with the same volume velocity: $\Re Z_A\to\rho ck^2/4\pi$ against $\rho ck^2/2\pi$. The cap radiates into
$4\pi$ and the piston into $2\pi$, so for a given volume velocity the cap radiates 3\,dB less
power than the infinite-baffle model predicts, which enters directly into the efficiency and
sensitivity of the loudspeaker;\citep{aarts2006} on axis this appears as the 6\,dB baffle step.
\item The low-frequency acoustic radiation mass of the uniform (rigid) cap with
$\theta_0=\pi/8$ is 0.758 times that of the piston ($\tfrac83\rho a_d^3$) for $K=0$, and 0.783, 0.801 and 0.813
times that of the Stenzel piston with the same $K$ for $K=1,2,3$.
\item The cap has essentially no resistance overshoot: for $K=0$ its maximum is 0.964,
against a high-frequency limit of 0.962, whereas the piston peaks at 1.13. The reactance of the
cap does not drop as deeply after its maximum and stays above that of the piston at high
$ka_d$. For the rigid cap the high-frequency resistance tends to
$2(1-x_0^3)/(3\sin^2\theta_0)=0.962$ instead of 1, because the normal velocity $v_0\cos\theta$
acts over the curved cap area, which is larger than $S_d$.
\end{enumerate}

\section{Conclusions}
The radiation impedance of an axially moving spherical cap on a rigid sphere, a simple model
of a closed-box loudspeaker, can be written exactly as a series connection of ideal
transformers and finite Cauer ladders, one per spherical mode. The element values are
positive, independent of frequency and given by Eq.~\eqref{eq:elements}. The transformer
ratios follow from the Legendre coefficients of the cone velocity. Truncating at $N$ modes and
adding a residual mass gives a passive, finite-dimensional implementation of the infinite modal
model, whose accuracy is controlled by $N$. The network
connects directly to the mechanical side of a loudspeaker equivalent circuit, and it can be
used in both frequency- and time-domain simulations. Despite the quite different geometry, the impedance of the cap resembles that of the
baffled piston: it behaves like the piston at high frequency and like a small closed box,
radiating into $4\pi$, at low frequency. Together with the earlier agreement in polar response,
baffle step, sound power and directivity, this supports the cap as a model of the radiation load
of a closed-box loudspeaker. Compared with the baffled-piston load at equal
volume velocity, the low-frequency resistance is halved, so that the radiated power is 3\,dB
lower, and for a rigid cap with $\theta_0=\pi/8$ the radiation mass is reduced by about a quarter;
how much this changes the efficiency and resonance frequency of a complete loudspeaker depends
on its other parameters.

\begin{acknowledgments}
The author used an AI language model to assist with editing the English text.
\end{acknowledgments}

\section*{Author Declarations}
\noindent\textit{Conflict of Interest.} The author has no conflicts to disclose.

\section*{Data Availability}
The data that support the findings of this study are available from the corresponding author
upon reasonable request.

\bibliography{capladder}

\begin{thebibliography}{10}
\def\enquote#1,{``#1,''}
\def\enxquote#1{``#1''}
\expandafter\ifx\csname url\endcsname\relax
  \def\url#1{\texttt{#1}}\fi
\expandafter\ifx\csname urlprefix\endcsname\relax\def\urlprefix{URL }\fi
\providecommand{\bibinfo}[2]{#2}
\def\plainquote#1{``#1''}
\providecommand{\noopsort}[1]{}
\providecommand{\switchargs}[2]{#2#1}
\providecommand{\dourl}[1]{\href{http://#1}{\nolinkurl{#1}}}
  \def\eatspace #1{#1}

\bibitem{beranek2012}
\bibinfo{author}{L.~L. Beranek} and \bibinfo{author}{T.~J. Mellow},
  \emph{\bibinfo{title}{Acoustics: Sound Fields and Transducers}}
  (\bibinfo{publisher}{Academic Press}, \bibinfo{address}{Oxford},
  \bibinfo{year}{2012}).

\bibitem{aj2003}
\bibinfo{author}{R.~M. Aarts} and \bibinfo{author}{A.~J. E.~M. Janssen},
  \enquote{\bibinfo{title}{Approximation of the {Struve} function
  {$\mathbf{H}_1$} occurring in impedance calculations}},  \bibinfo{journal}{J.
  Acoust. Soc. Am.} \textbf{113}, \bibinfo{pages}{2635--2637}
  (\bibinfo{year}{2003}).

\bibitem{aj2016}
\bibinfo{author}{R.~M. Aarts} and \bibinfo{author}{A.~J. E.~M. Janssen},
  \enquote{\bibinfo{title}{Efficient approximation of the {Struve} functions
  {$\mathbf{H}_n$} occurring in the calculation of sound radiation
  quantities}},  \bibinfo{journal}{J. Acoust. Soc. Am.} \textbf{140},
  \bibinfo{pages}{4154--4160} (\bibinfo{year}{2016}).

\bibitem{morse1968}
\bibinfo{author}{P.~M. Morse} and \bibinfo{author}{K.~U. Ingard},
  \emph{\bibinfo{title}{Theoretical Acoustics}}
  (\bibinfo{publisher}{McGraw-Hill}, \bibinfo{address}{New York},
  \bibinfo{year}{1968}) \bibinfo{note}{{Chap.}~7}.

\bibitem{aj2010}
\bibinfo{author}{R.~M. Aarts} and \bibinfo{author}{A.~J. E.~M. Janssen},
  \enquote{\bibinfo{title}{Sound radiation from a resilient spherical cap on a
  rigid sphere}},  \bibinfo{journal}{J. Acoust. Soc. Am.} \textbf{127},
  \bibinfo{pages}{2262--2273} (\bibinfo{year}{2010}).

\bibitem{aj2011}
\bibinfo{author}{R.~M. Aarts} and \bibinfo{author}{A.~J. E.~M. Janssen},
  \enquote{\bibinfo{title}{Comparing sound radiation from a loudspeaker with
  that from a flexible spherical cap on a rigid sphere}},  \bibinfo{journal}{J.
  Audio Eng. Soc.} \textbf{59}, \bibinfo{pages}{201--212}
  (\bibinfo{year}{2011}).

\bibitem{sherman1959}
\bibinfo{author}{C.~H. Sherman}, \enquote{\bibinfo{title}{Mutual radiation
  impedance of sources on a sphere}},  \bibinfo{journal}{J. Acoust. Soc. Am.}
  \textbf{31}, \bibinfo{pages}{947--952} (\bibinfo{year}{1959}).

\bibitem{bellows2023}
\bibinfo{author}{S.~D. Bellows} and \bibinfo{author}{T.~W. Leishman},
  \enquote{\bibinfo{title}{Low-frequency radiation from a vibrating cap on a
  rigid spherical shell with a circular aperture}},  \bibinfo{journal}{J.
  Acoust. Soc. Am.} \textbf{154}, \bibinfo{pages}{3883--3898}
  (\bibinfo{year}{2023}).

\bibitem{bellows2024}
\bibinfo{author}{S.~D. Bellows}, \enquote{\bibinfo{title}{Comparing the end
  correction of a spherically baffled piston to infinitely baffled and
  unbaffled circular radiators {(L)}}},  \bibinfo{journal}{J. Acoust. Soc. Am.}
  \textbf{155}, \bibinfo{pages}{3302--3305} (\bibinfo{year}{2024}).

\bibitem{helie2003}
\bibinfo{author}{T.~H\'elie} and \bibinfo{author}{X.~Rodet},
  \enquote{\bibinfo{title}{Radiation of a pulsating portion of a sphere:
  Application to horn radiation}},  \bibinfo{journal}{Acta Acust. united
  Acust.} \textbf{89}, \bibinfo{pages}{565--577} (\bibinfo{year}{2003}).

\bibitem{guillemin1957}
\bibinfo{author}{E.~A. Guillemin}, \emph{\bibinfo{title}{Synthesis of Passive
  Networks}}  (\bibinfo{publisher}{Wiley}, \bibinfo{address}{New York},
  \bibinfo{year}{1957}).

\bibitem{chu1948}
\bibinfo{author}{L.~J. Chu}, \enquote{\bibinfo{title}{Physical limitations of
  omni-directional antennas}},  \bibinfo{journal}{J. Appl. Phys.} \textbf{19},
  \bibinfo{pages}{1163--1175} (\bibinfo{year}{1948}).

\bibitem{hansen1981}
\bibinfo{author}{R.~C. Hansen}, \enquote{\bibinfo{title}{Fundamental
  limitations in antennas}},  \bibinfo{journal}{Proc. IEEE} \textbf{69},
  \bibinfo{pages}{170--182} (\bibinfo{year}{1981}).

\bibitem{aarts2026}
\bibinfo{author}{R.~M. Aarts}, \enquote{\bibinfo{title}{Continued fractions and
  passive ladder networks for acoustic impedances: the baffled piston and the
  horn}},  \bibinfo{journal}{arXiv:2610.20371, October 14, 2026, preprint, v1}
  .

\bibitem{kjerbye1956}
\bibinfo{author}{A.~{Kjerbye Nielsen}}, \enquote{\bibinfo{title}{Equivalent
  network models of continuous media and a network representation of sound
  radiators}},  \bibinfo{journal}{J. Acoust. Soc. Am.} \textbf{28},
  \bibinfo{pages}{788} (\bibinfo{year}{1956}) \bibinfo{note}{(abstract)}.

\bibitem{olver2010}
\bibinfo{editor}{F.~W.~J. Olver}, \bibinfo{editor}{D.~W. Lozier},
  \bibinfo{editor}{R.~F. Boisvert}, and \bibinfo{editor}{C.~W. Clark}, eds.,
  \emph{\bibinfo{title}{{NIST} Handbook of Mathematical Functions}}
  (\bibinfo{publisher}{Cambridge University Press},
  \bibinfo{address}{Cambridge}, \bibinfo{year}{2010}).

\bibitem{aartsproof}
\bibinfo{author}{R.~M. Aarts}, \enquote{\bibinfo{title}{The ladder network of
  the baffled-piston radiation impedance: an elementary proof of existence to
  all orders and of the limiting behaviour of the ladder coefficients}},
  \bibinfo{journal}{arXiv, November 10, 2026, preprint, v1} .

\bibitem{aj2009}
\bibinfo{author}{R.~M. Aarts} and \bibinfo{author}{A.~J. E.~M. Janssen},
  \enquote{\bibinfo{title}{Sound radiation quantities arising from a resilient
  circular radiator}},  \bibinfo{journal}{J. Acoust. Soc. Am.} \textbf{126},
  \bibinfo{pages}{1776--1787} (\bibinfo{year}{2009}).

\bibitem{aj2012}
\bibinfo{author}{R.~M. Aarts} and \bibinfo{author}{A.~J. E.~M. Janssen},
  \enquote{\bibinfo{title}{Sound radiation from a loudspeaker, from a spherical
  pole cap, and from a piston in an infinite baffle}},  \bibinfo{journal}{Noise
  \& Vibration Worldwide} \textbf{43}, \bibinfo{pages}{12--19}
  (\bibinfo{year}{2012}).

\bibitem{aarts2006}
\bibinfo{author}{R.~M. Aarts}, \enquote{\bibinfo{title}{Optimally sensitive and
  efficient compact loudspeakers}},  \bibinfo{journal}{J. Acoust. Soc. Am.}
  \textbf{119}, \bibinfo{pages}{890--896} (\bibinfo{year}{2006}).

\end{thebibliography}

\end{document}